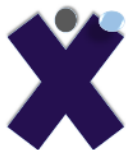

# Mathematics and Sports

# An Analysis of an Alternative Pythagorean Expected Win Percentage Model: Applications Using Major League Baseball Team Quality Simulations

Justin Ehrlich[1,*], Christopher Boudreaux[2], James Boudreau[3] and Shane Sanders[1]

[1] Sport Analytics, Syracuse University, Syracuse, NY, [2] Economics, Florida Atlantic University, Boca Raton, FL, [3] Economics, Kennesaw State University, Kennesaw, GA

*Corresponding Author E-mail:
jaehrlic@syr.edu

**Abstract**

*Background*. Contests are games in which the players compete for a prize and exert effort to increase their probability of winning. For sport contests, analysts often use the Pythagorean expected win percentage model to estimate teams' expected wins (quality). We ask if there are alternative contest models that minimize error or information loss from misspecification and outperform the Pythagorean model.

*Aim*. This article aims to use simulated data to select the optimal expected win percentage model among the choice of relevant alternatives. The choices include the traditional Pythagorean model and the difference-form contest success function (CSF).

*Method*. We simulate 1,000 iterations of the 2014 MLB season for the purpose of estimating and analyzing alternative models of expected win percentage (team quality). We use the open-source, Strategic Baseball Simulator and develop an AutoHotKey script that programmatically executes the SBS application, chooses the correct settings for the 2014 season, enters a unique ID for the simulation data file, and iterates these steps 1,000 times. We estimate expected win percentage using the traditional Pythagorean model, as well as the difference-form CSF model that is used in game theory and public choice economics. Each model is estimated while accounting for fixed (team) effects.

*Result*. We find that the difference-form CSF model outperforms the traditional Pythagorean model in terms of explanatory power and in terms of misspecification-based information loss as estimated by the Akaike Information Criterion. Through parametric estimation, we further confirm that the simulator yields realistic statistical outcomes.

*Conclusion*. The simulation methodology offers the advantage of greatly improved sample size. As the season is held constant, our simulation-based statistical inference also allows for estimation and model comparison without the (time series) issue of non-stationarity. The results suggest that improved win (productivity) estimation can be achieved through alternative CSF specifications.

## 1 Introduction

The Pythagorean expected win percentage model was developed by Bill James (1980) to estimate a baseball team's expected win percentage, as distinct from its actual wins, over the course of a season. As such, the model can be used to assess team luck or misfortune (actual wins – expected wins). Such knowledge is valuable, as future outcomes often depend upon expected, rather than realized, outcomes (Leung and Joseph, 2014). From a managerial perspective, measuring and analyzing errors in forecasting over many different possible scenarios provides better measures of managerial ability (Teach, 1990). Simulation-based statistical inference, therefore, highlights the importance of understanding how reproducible a given result may be in the next time period (Winston, 2012), which is an emerging area of importance in interdisciplinary research fields such as data science, statistics, economics, and management (Kamihigashi, 2018). Moreover, a process-oriented organization may wish to reward employees based upon expected rather than realized outcomes. The expectancy theory of motivation (Vroom, 1964) highlights how employees exert effort according to their expectations about the rewards associated with that effort. To be certain, realized outcomes matter, but over time, rational expectations will be adjusted to closely match realized outcomes (Muth, 1961).

The Pythagorean expected win percentage model in baseball (and subsequently other sports) is an example of what social scientists refer to as a Tullock (1980) or additive-form contest success function (CSF). The model is used widely because it is straightforward (e.g., highly accessible for baseball scouts and other practitioners) and yet historically explains a high proportion of variation in wins (see, e.g., Kovalchik, 2016). Herein, we test whether the model represents an optimal specification among the set of standard contest success functional forms by which to estimate expected win percentage in major league baseball. Using the *Strategic Baseball Simulator* (SBS), a leading, event-level Major League Baseball simulation software, we simulate 1,000 iterations of the 2014 MLB regular season data to estimate team win proportions using Tullock-form and difference-form CSF models.[1][2]

In terms of testing the expected win percentage models, the simulation-based approach has clear advantages. Firstly, we are able to generate 2.43 million simulated MLB game outcomes, more than 10 times as many game outcomes as have ever been generated empirically. Moreover, the simulation-based approach avoids the problem of time series non-stationarity, which may create difficulty in estimating a best fit expected win percentage model. To collect a large sample of game outcomes, the empirical approach usually involves collection of data across several years. However, the level of noise in the mapping from game inputs to game outputs can change structurally from year-to-year such that this approach is never directly estimating an optimized Pythagorean expected win percentage model for any particular season. Indeed, the simulation-based approach has proven valuable in generating large samples of simulated events to better understand complex phenomena at a given event in a specified time period (see, e.g., Goto et al., 2019 to understand the value of simulation in predicting economic performance and outcomes; see, e.g., Ozaki et al., 2019 for an examination as to the economic and financial effects of natural disaster). There is a limitation of the simulation-based approach in that a simulated season is simply a model of a season and not an actual season. By assessing secondary characteristics of the simulated data results relative to those of known empirical data results, however, we are able to assess the real-world validity of the SBS-generated data set. In fact, we find that the SBS-generated data set aligns closely to real-world data with respect to expected win percentage estimation in MLB.

Applying the Akaike (1981) Information Criterion for model specification to the simulated data, we find evidence that the difference form CSF minimizes misspecification associated information loss. That is to say, there is evidence that the originally-specified Tullock-form Pythagorean expected win percentage model is not an optimal CSF specification by which to estimate expected win percentage. Following James (1980), there is a vast literature on the topic of Pythagorean expected win percentage (see, e.g., recent contributions by Chen and Li 2016, Kovalchik 2016, Vine 2016, and Caro and Machtmes 2013). These works estimate expected win percentage using empirical sports data. The present contribution is unique in its use of simulated data to consider the specification of the Pythagorean expected win percentage model.

## 2  Methods

To simulate 1,000 iterations of the 2014 MLB season, we used Strategic Baseball Simulator (SBS), which is a well-known and well-regarded MLB game and season simulator. SBS is open source and is based entirely on roster statistics to simulate games. By feeding SBS roster data, a user can simulate a game between any two teams. SBS also allows a user to simulate and output multiple iterations of a 162-game season for each of the 30 competing MLB teams. We provided SBS with the 2014 MLB season and simulated the event 1,000 times. For each iteration, we collected the results in a unique file. We then developed a Java application to read each file, acquire the pertinent information about each season, and output the results to a comma-separated value (CSV) file. As running SBS manually 1,000 times would require a substantial amount of time, we decided to use an open source scripting language called AutoHotKey to automate the work. We wrote an AutoHotKey script that programmatically executes the SBS application, chooses the correct settings for the 2014 season, enters a unique ID for the simulation data file, and repeats these steps 1,000 times.

## 3  Results and Discussion

The Pythagorean estimated wins model was created by pioneering sabermetrician Bill James (1980). In its original form, the model appeared as follows:

---

[1] The 2014 season was selected because it was the latest season for which the SBS application features all data needed to run the simulation.
[2] For a comparison of these CSF forms, see, e.g., Hirshleifer (1989).

$$\text{Expected win percentage}_{i,j} = \frac{rs_{i,j}^2}{rs_{i,j}^2 + ra_{i,j}^2} \quad (1)$$

where $rs_{i,j}$ represents runs scored by team i in season j, $ra_{i,j}$ represents runs allowed by team i in season j, and the original model restricts the noise parameter (in the exponent of each argument) to the value 2. In its general form, the Pythagorean expected win percentage model appears as in (2).

$$\text{Expected win percentage}_{i,j} = \frac{rs_{i,j}^\alpha}{rs_{i,j}^\alpha + ra_{i,j}^\alpha} \quad (2)$$

where $\alpha(>0)$ is defined in contest theory as the "noise" or level of determinism in the mapping from inputs (runs scored and allowed) to output (wins) (see, e.g., Nti, 1999). The degree of contest noise decreases in α. As stated in the introduction, the Pythagorean expected win percentage model is an example of a Tullock-form contest success function (Tullock, 1980). We also consider the difference form contest success function as a model of expected win percentage estimation:

$$\text{Expected win percentage}_{i,j} = \frac{1}{1 + e^{\alpha(ra_{i,j} - rs_{i,j})}} \quad (3)$$

A difference-form contest success function takes a logistic functional form. As a teaching note, logistic regression cannot be used, however, as the left-hand side variable denotes winning proportion in a season, which is non-binary. We therefore log-transform each respective model into one that is linear in the parameters and subsequently run OLS regression. Table 1 provides the results for Tullock-form and difference-form CSFs.

**Table 1**. Expected Win Percentage Models for Tullock-Form and Difference-Form CSFs

|  | Contest Success Functions (CSF) | |
|---|---|---|
|  | Tullock-Form (1) | Difference-Form (2) |
| Estimated parameter | 1.722**** (0.005) | 0.003**** (0.000006) |
| Number of observations | 1,000 | 1,000 |
| $R^2$ | 0.826 | 0.829 |
| Akaike Information Criterion (AIC) | -49,671.95 | -50,128.41 |
| Root Mean Squared Error (MSE) | 0.106 | 0.105 |

*Note* – Standard errors in parentheses. ****$p < 0.001$ (two-tailed t-test).

Column 1 of Table 1 reports results of the Tullock-form model, which has an estimated coefficient value of 1.72.[3] This estimate is in line with prior empirical estimates for the Tullock-form, Pythagorean expected win percentage model (see, e.g., Dayaratna and Miller, 2012). We also observe that this model has a fairly high R-squared value (0.826), consistent with an empirically-estimated Pythagorean model. This provides additional evidence that the simulated data-generating process is consistent with the fundamental empirical data-generating process that it seeks to simulate. According to the Tullock-form model results, then, an optimized Tullock-form expected win percentage model given the simulation data appears as follows:

$$\text{Expected win percentage}_{i,j} = \frac{rs_{i,j}^{1.72}}{rs_{i,j}^{1.72} + ra_{i,j}^{1.72}} \quad (4)$$

The exponent in the Tullock-form is a returns to scale parameter (see, e.g., Nti, 1999). This exponent indicates the degree of noise in the contest, where smaller values indicate a noisier contest. Therefore, our simulation sample indicates a slightly noisier mapping from runs scored and runs allowed to win proportion than that originally envisioned by Bill James (1980). We now consider the difference-form CSF wins estimation model.

Column 2 of Table 1 reports the difference-form CSF expected win percentage model. Within the Pythagorean expected win percentage literature, there are no prior expected win percentage estimations under the difference-form specification. In other words, this is a novel application of the difference-form CSF, and there is no prior evidence that it should perform better (worse) than the traditional Pythagorean expected win percentage model. In fact, we find that the difference-form model has a slightly higher R-squared value than does the standard (Tullock-form) Pythagorean expected win percentage model. That is to say, runs scored and runs allowed explain a higher proportion of variation in win proportion within the difference-form specification. To get a sense as to the statistical magnitude of this difference, we generate post-estimation Akaike Information Criterion (AIC) values for each respective regression. To do so, we used the *estat ic* command within Stata. These values provide an estimate as to the misspecification-based information loss for each model, where a lower value indicates less information loss.

---

[3] In past estimations of the Pythagorean expected wins model, it has been common to round the optimized value of the exponent to the nearest hundredth when the model is used (see, e.g., Dayaratna and Miller, 2012).

The AIC values are qualitatively consistent with the R-squared values. They indicate that the difference-form minimizes misspecification-based information loss among the two specifications (i.e., that the difference-form is estimated to represent a better specification given the data). To get a sense of the magnitude of difference in information loss, we apply the following formula:

$$\frac{P_T}{P_D} = e^{\frac{-50{,}128.41-(-49{,}671.95)}{2}} = e^{-228.23} \approx 7.60 \times 10^{-100} \tag{5}$$

where $\frac{P_T}{P_D}$ represents the estimated likelihood that the Tullock-form minimizes misspecification-based information loss ($P_T$) relative to the likelihood that the difference-form does this ($P_D$). In other words, it is exceedingly unlikely, according to the AIC ratio, that the Tullock-form represents a best fit of the data. This result provides evidence that the difference-form allows for a better-specified wins estimation model than the traditionally-used Tullock-form (see, e.g., Hu, 2007 for more information on interpreting AIC results). Therefore, the difference form CSF model we estimate is the following:

$$\text{Expected win percentage}_{i,j} = \frac{1}{1 + e^{0.003(ra_{i,j} - rs_{i,j})}} \tag{6}$$

This is the optimized difference-form model based upon an OLS estimation of the simulation data generated. We note that the parametric estimate differs by orders of magnitude due to underlying differences between the distribution of runs scored differences by team and that of runs scored ratio by team.

## 4  Conclusion

We used Strategic Baseball Simulator (SBS) to simulate 1,000 iterations of the 2014 MLB season. For each iteration, we collected the results in a unique file. We then developed a Java application to read each file, acquire the pertinent information about each season, and output the results to a comma-separated value (CSV) file. As running SBS manually 1,000 times would require a substantial amount of time, we decided to use an open source scripting language, called AutoHotKey, to automate the process. We wrote an AutoHotKey script that programmatically executes the SBS application, chooses the correct settings for the 2014 season, enters a unique ID for the simulation data file, and repeats these steps 1,000 times.

Using the simulated data, we estimated the traditional Pythagorean model of expected win percentage, as well as an alternative model of expected win percentage in which a difference-form contest success function is specified. As discussed, the simulation approach has strong advantages in that a) we were able to generate more than ten times as much MLB season data as has ever been generated in real play, and b) the simulation-based approach avoids the problem of time series non-stationarity (as may be present in any empirical analysis). Our simulation results suggest that the proposed difference-form model improves upon the traditional Pythagorean model in terms of model fit (according to both the R-squared and Akaike Information Criterion). We conclude that the traditional form of the Pythagorean model is not optimized with respect to either explanatory power and misspecification-based information loss. With an equally simplistic model, we can more closely predict and estimate expected win percentage.

## Conflict of Interest Statement

The authors declare they have no conflict of interest and no financial motive in this study.

## Data availability statement